\documentclass[a4paper, 12pt]{article}

\usepackage[T1]{fontenc}
\usepackage{lmodern}

\usepackage{setspace}
\usepackage{pdflscape}
\usepackage{longtable}
\usepackage{wasysym}
\usepackage{hyperref}
\usepackage{booktabs} 
\usepackage{tabularx} 
\usepackage{graphicx} 
\usepackage{array}    
\hypersetup{
    colorlinks=true,
    linkcolor=blue,
    urlcolor=blue,
    citecolor=blue}
\usepackage{pdflscape}
\usepackage{comment}

\begin{document}
\begin{center}


\textbf{\Large Ten Years of the Stochastic Resonance Model of Tinnitus: From Phantom Perception to Adaptive Sensory Optimization}
\newline

{\sc Patrick Krauss$^{1,2,3,4}$, Achim Schilling$^{3,4,5}$}\\
\vspace{0.2cm}
{\small $^1$ Cognitive Computational Neuroscience Group, Pattern Recognition Lab, Friedrich-Alexander-University Erlangen-Nürnberg (FAU), Germany\\
$^2$ Physics and Cognition Group, Mannheim Center for Neuromodulation and Neuroprosthetics (MCNN), University Hospital Mannheim, University Heidelberg, Germany\\
$^3$ Neuroscience Lab, University Hospital Erlangen, Germany\\
$^4$ BG Clinic Ludwigshafen, Germany\\
$^5$ NeuroAI and BCI Group, Mannheim Center for Neuromodulation and Neuroprosthetics (MCNN), University Hospital Mannheim, University Heidelberg, Germany}

\end{center}

\vspace{1cm}


\begin{abstract}
\noindent Subjective tinnitus - the perception of sound in the absence of an external acoustic stimulus - remains one of the most debated phenomena in auditory neuroscience. In 2016, the stochastic resonance (SR) model was introduced as an alternative account of tinnitus-related neuronal hyperactivity, proposing that internally generated neural noise is adaptively upregulated to restore information transmission after hearing loss. Rather than interpreting increased spontaneous activity as purely maladaptive, the model reframed it as a functional mechanism that enhances signal detection near sensory thresholds, with tinnitus emerging as a perceptual side effect of adaptive sensory optimization.
Over the past decade, this framework has evolved from a phenomenological hypothesis into a broader neurocomputational theory integrating information theory, adaptive signal detection, multichannel auditory processing, and cross-modal plasticity. Computational simulations, large-scale clinical data analyses, and experimental findings from animal studies have provided converging support for key predictions of the model, including improved detectability under specific noise regimes and frequency-specific phantom percepts. The SR framework has further been extended into a therapeutic approach using spectrally matched near-threshold noise stimulation and has recently been embedded within a unified theory of auditory (phantom) perception that integrates stochastic resonance, central gain, homeostatic plasticity, and predictive coding.
This review provides a chronological account of the development of the stochastic resonance model of tinnitus, summarizes theoretical refinements and empirical evidence, and outlines future directions for mechanistic validation and clinical translation. By redefining tinnitus as an emergent consequence of adaptive sensory computation, the model shifts the conceptual focus from pathological dysfunction to principles of information optimization in neural systems.
\end{abstract}

\newpage


\section{Introduction}

Tinnitus affects approximately $10$--$15\%$ of the population and constitutes one of the most prevalent examples of a sensory phantom percept \cite{henry2020tinnitus}. Despite extensive experimental and theoretical work, a unified mechanistic account of tinnitus generation has not yet emerged. A large body of research links tinnitus to peripheral hearing loss and subsequent changes in central auditory processing, most prominently increased spontaneous firing rates and neural hyperactivity along the auditory pathway \cite{kaltenbach1996increases, ma2006spontaneous, norena2003changes, ropp2014effects}. Classical explanatory frameworks, including central gain and homeostatic plasticity models, interpret these changes as compensatory responses to reduced sensory input that become maladaptive over time \cite{auerbach2014central, roberts2019overview, yang2011homeostatic}. While such models successfully account for several physiological and behavioral observations, neuronal hyperactivity is typically regarded as an epiphenomenon or pathological side effect of compensation rather than as a functional component of auditory processing.

In 2016, the stochastic resonance (SR) model introduced an alternative conceptual perspective by proposing that increased spontaneous activity may serve a functional role in sensory processing. In this framework, internally generated noise is adaptively regulated to improve signal detection and maintain optimal information transmission under conditions of degraded input \cite{krauss2016stochastic}. From this viewpoint, tinnitus-related hyperactivity does not primarily reflect dysfunctional amplification but may instead arise as a consequence of an adaptive neural mechanism operating near sensory detection thresholds. In the model, tinnitus is therefore interpreted not as the primary failure of the auditory system, but as a perceptual side effect of a compensatory process aimed at restoring sensitivity following hearing loss. In this sense, tinnitus-related hyperactivity is interpreted as a functional consequence of adaptive sensory optimization aimed at maintaining information transmission under degraded input conditions.

The present article provides a chronological review of the development of the stochastic resonance model over the past decade. We trace its conceptual origins, summarize key theoretical and computational advances, and discuss how its predictions relate to empirical findings in auditory neuroscience. By situating the SR framework within the broader landscape of tinnitus models, this review aims to clarify both its explanatory scope and its limitations, and to outline open questions for future research.

\section{Existing Models and open questions}
The following section provides a brief introduction to the most prominent models of tinnitus development. This overview is not intended to be exhaustive, but rather to highlight the most influential models that have shaped the field. Although these approaches have substantially advanced our understanding of neural adaptations following hearing loss, significant questions remain unanswered.

\subsection{Homeostatic Plasticity Models}
Homeostatic plasticity models propose that tinnitus arises from regulatory mechanisms that stabilize neuronal firing rates in response to reduced peripheral input \cite{yang2011homeostatic}. Reduced auditory nerve activity triggers compensatory increases in synaptic gain or intrinsic excitability to restore baseline firing levels. These mechanisms are well supported by experimental evidence and explain the emergence of elevated spontaneous firing rates after hearing loss.
Within this framework, tinnitus-related hyperactivity is generally considered a harmful byproduct of beneficial stabilization processes. Homeostatic plasticity models primarily address adaptations that occur over the course of hours to weeks \cite{zenke2017temporal}, providing a plausible explanation for persistent central changes.

\subsection{Central Gain Models}
Central gain models emphasize the amplification of neural responses along the auditory pathway due to reduced sensory input. While increased gain compensates for diminished peripheral input, it can also produce excessive spontaneous or synchronized activity, which is perceived as tinnitus \cite{auerbach2014central, roberts2019overview}. These models explain the link between hearing loss, hyperacusis, and increased neural activity in brainstem and cortical regions.
Conceptually, central gain may not be a fundamentally different mechanism than homeostatic plasticity; rather, it may describe similar compensatory processes at a coarser systems level \cite{schilling2023predictive}. In this view, changes in gain could reflect the aggregate outcome of underlying homeostatic adjustments in synaptic strength or intrinsic excitability.
Similar to homeostatic plasticity models, increased activity is generally interpreted as pathological overcompensation rather than a functional mechanism that improves signal detection directly. 

\subsection{Predictive Coding and Bayesian Inference Models}
Predictive coding and Bayesian inference frameworks interpret tinnitus as a consequence of perceptual inference \cite{sedley2016integrative, hullfish2019prediction, de2024bayesian}. Perception emerges from the interaction between sensory evidence and internally generated predictions. When sensory input becomes unreliable, internally generated expectations may dominate perception, leading to phantom percepts.
These models explain the influence of attention, expectation, world models and contextual factors on tinnitus. However, they focus primarily on higher level inferential processes and typically interpret tinnitus as misattributed internally generated activity rather than as a direct consequence of altered early sensory processing.

\subsection{Open Questions}
Despite their strengths, several phenomena remain insufficiently explained.
Somatic tinnitus, for example, challenges models based solely on auditory gain or homeostatic plasticity, as these mechanisms operate on relatively slow timescales, whereas the tinnitus percept can be modulated almost instantaneously by jaw or neck movements \cite{shore2007neural}. Similarly, the near-immediate onset of tinnitus after acoustic trauma or during exposure to anechoic conditions is difficult to reconcile with models based on homeostatic plasticity or gain changes that typically emerge only over much longer timescales \cite{fagelson2007association, almond2013transient}.
Perhaps the most compelling example is the Zwicker tone, a transient phantom percept induced by spectrally notched noise, which emerges within seconds despite the absence of structural hearing loss, thereby challenging models that rely on slow adaptive processes \cite{zwicker1964negative, parra2007illusory}.
Finally, although Bayesian frameworks provide a convincing computational description, it is unclear how predictive coding principles are mechanistically implemented in the specific neural circuits of the auditory system \cite{sedley2016integrative, de2024bayesian}.
The stochastic resonance model was developed to address these open questions. Unlike existing accounts, the model proposes a concrete functional mechanism operating on short timescales and suggests specific neural substrates through which adaptive internal noise regulation may cause tinnitus or Zwicker tone as a byproduct of improved signal detection \cite{schilling2020stochastic}.

\section{History of the Stochastic Resonance Model of Tinnitus}
The stochastic resonance model of tinnitus did not emerge in isolation but developed at the intersection of information theory, auditory neuroscience, and the broader study of noise-enhanced signal processing in biological systems. Over the past decade, the framework evolved from an initial hypothesis explaining tinnitus-related hyperactivity after hearing loss into a broader theory of adaptive sensory processing, phantom perception, and clinical intervention. The following sections trace this development chronologically, highlighting key conceptual advances, empirical findings, and theoretical extensions.

\subsection{Before 2016: Stochastic Resonance, Noise and Information Theory}

Prior to its application to tinnitus, stochastic resonance (SR) had been established as a general phenomenon in nonlinear dynamical systems \cite{benzi1981mechanism, gammaitoni1998stochastic}. Initially described in physical contexts, SR demonstrated that the addition of noise to a nonlinear detector can improve signal transmission when inputs fall below detection threshold. Rather than degrading performance, noise under appropriate conditions can enhance the detectability of weak signals by enabling threshold crossings that would otherwise not occur. This was, for example, demonstrated by adding mild noise to acoustic stimuli below the detection threshold \cite{zeng2000human}. Subsequent experimental and theoretical work extended this principle to biological systems, showing that noise-enhanced signal detection can arise in sensory receptors and neural populations, thereby establishing SR as a biologically plausible mechanism rather than a purely physical curiosity \cite{itzcovich2017stochastic, moss2004stochastic, manjarrez2003stochastic}. These developments provided the theoretical foundation for later applications of SR to auditory processing, tinnitus, and phantom perception.

\subsection{2016: The First Formulation of the Stochastic Resonance Model of Tinnitus}
The stochastic resonance model of tinnitus was first formally introduced in 2016 by Krauss and colleagues, who proposed that tinnitus-related neuronal hyperactivity may arise from an adaptive increase of internally generated noise following hearing loss \cite{krauss2016stochastic}. The model was motivated by the observation that reduced peripheral input decreases the detectability of weak sensory signals, while increased spontaneous activity is consistently observed at multiple levels of the auditory pathway in tinnitus \cite{kaltenbach1996increases, ma2006spontaneous, norena2003changes, ropp2014effects}. Rather than interpreting this hyperactivity as purely maladaptive, the model suggested that it may reflect a compensatory mechanism aimed at preserving information transmission under degraded input conditions.

\subsubsection{Conceptual Innovation}
The central conceptual contribution of the model consisted of two closely related shifts in interpretation. First, spontaneous neural activity was reinterpreted functionally. Increased spontaneous firing rates were not assumed to represent pathological overexcitation, but were instead proposed to constitute internally generated noise required for optimal signal detection in nonlinear sensory systems. Within this framework, neuronal hyperactivity emerges as a necessary by-product of an adaptive process operating near detection threshold \cite{krauss2016stochastic}.

Second, the model proposed that the auditory system regulates internal noise levels so as to maximize information transmission between sensory input and neural response. This information-theoretic perspective later motivated the development of biologically plausible adaptive stochastic resonance mechanisms based on output autocorrelation \cite{krauss2017adaptive}.

This interpretation differs fundamentally from homeostatic plasticity and central gain models \cite{auerbach2014central, roberts2019overview, yang2011homeostatic}. Whereas these frameworks assume compensatory gain changes acting on already detected signals, the stochastic resonance model operates at the level of signal detection itself. By increasing internal noise, subthreshold inputs can cross neuronal thresholds, thereby improving detectability before amplification mechanisms become relevant.

\subsubsection{Computational Implementation}
The original model was formulated as a phenomenological feedback system consisting of three functional components: a sensory detector representing the cochlear or early auditory stage, an information detector computing the autocorrelation of neural responses, and a noise generator whose output was regulated through feedback \cite{krauss2016stochastic}. Reduced sensory input following hearing loss decreased output autocorrelation, which in turn increased internally generated noise until information transmission was partially restored.
Computational simulations demonstrated that this adaptive noise regulation improves detection performance under elevated thresholds while simultaneously increasing spontaneous activity within the system \cite{krauss2016stochastic}. Within the proposed framework, this increased activity represents the neural correlate of tinnitus, emerging as a consequence of adaptive sensory optimization rather than as a primary dysfunction.

\subsubsection{Empirical Consistency}
In addition to its theoretical and computational formulation, the 2016 study analyzed audiometric data from over 39,000 patients \cite{krauss2016stochastic, gollnast2017analysis}. Tinnitus patients exhibited improved low-frequency hearing thresholds compared to patients without tinnitus, which is consistent with threshold compensation mediated by stochastic resonance \cite{gollnast2017analysis}.
Subsequent work focused on biological plausibility by linking the functional components of the stochastic resonance framework to auditory structures known to be capable of implementing adaptive noise regulation. Rather than assigning a single anatomical location, stochastic resonance was interpreted as a functional principle operating at multiple early processing stages \cite{krauss2016stochastic}. Particular attention was given to the lateral olivocochlear efferent system, which influences auditory nerve activity as a descending pathway, and to the dorsal cochlear nucleus, whose circuitry supports the processing and integration of temporal correlations and auditory and somatosensory inputs \cite{krauss2016stochastic}. The dorsal cochlear nucleus provides a mechanistic link to somatic tinnitus and exhibits rapid activity changes after acoustic trauma, which is consistent with adaptive processes operating on short timescales \cite{krauss2016stochastic, shore2005multisensory, shore2006somatosensory}.

\subsubsection{Integration with Hidden Hearing Loss and Synaptopathy}
The SR framework also naturally accounts for tinnitus in individuals with clinically normal audiograms \cite{schaette2011tinnitus}. Hidden hearing loss and cochlear synaptopathy can reduce the fidelity of neural information transmission without substantially affecting audiometric thresholds. Because adaptive stochastic resonance responds to diminished information transfer rather than absolute hearing thresholds, the model predicts compensatory upregulation of internal neural noise even in the absence of measurable hearing loss. This provides a mechanistic explanation for the frequent occurrence of tinnitus in patients whose standard audiometric measures remain within normal limits \cite{schilling2020stochastic}.

\subsubsection{Support from Ear-Plugging Studies}
Additional support for the stochastic resonance framework comes from ear-plugging studies in humans. In these experiments, temporary unilateral auditory deprivation is induced by inserting an ear plug into one ear, thereby reducing auditory input without causing structural damage to the auditory system. Remarkably, healthy participants frequently report the emergence of transient tinnitus-like phantom percepts during or shortly after the deprivation period \cite{schaette2012reversible, fournier2014loudness}. These findings are difficult to explain solely by permanent cochlear pathology because no lasting peripheral damage is present. Instead, they suggest that reduced auditory input alone can trigger neural adaptations leading to phantom sound perception. Within the stochastic resonance framework, ear plugging decreases information transmission from the affected ear, resulting in adaptive upregulation of internally generated neural noise to maintain sensitivity. The transient tinnitus percept is therefore interpreted as a direct consequence of this compensatory process. Ear-plugging studies thus provide an important experimental bridge between chronic tinnitus associated with hearing loss and transient phantom percepts such as the Zwicker tone, demonstrating that reduced sensory input can be sufficient to induce tinnitus-like sensations even in otherwise healthy auditory systems.

\subsubsection{Comparison with Competing Models}
Over the past decade, competing models of tinnitus have differed primarily in their proposed mechanisms and interpretations of neuronal hyperactivity. Homeostatic plasticity and central gain accounts attribute increased activity to compensatory amplification following reduced peripheral input. These accounts view tinnitus as a maladaptive byproduct of this upregulation \cite{auerbach2014central, roberts2019overview, yang2011homeostatic}. In contrast, predictive coding frameworks emphasize altered perceptual inference. These frameworks propose that unreliable sensory input shifts the balance toward internally generated predictions that become consciously perceived \cite{sedley2016integrative, hullfish2019prediction, de2024bayesian}. The stochastic resonance model, in contrast, interprets hyperactivity as the functional correlate of adaptive internal noise regulation that enhances signal detectability under degraded input conditions \cite{krauss2016stochastic}. According to this model, tinnitus emerges as an unavoidable side effect of this optimization process rather than as mere pathological overexcitation \cite{krauss2016stochastic}.

\subsection{2017--2022: From a Tinnitus Model to a General Theory of Adaptive Sensory Processing and Phantom Perception}

Evidence for the biological plausibility of the SR framework was strengthened in 2017 by the introduction of an adaptive stochastic resonance mechanism capable of regulating internal noise levels without requiring knowledge of the sensory input signal. Instead, the model demonstrated that output autocorrelation can serve as a biologically plausible proxy for information transmission and can be used to continuously adjust noise levels through a feedback loop, thereby maximizing signal detectability under unknown and variable input conditions. Importantly, the study showed that adaptive stochastic resonance is not restricted to Gaussian noise, periodic signals, or hard detection thresholds, as often assumed, but remains effective for non-Gaussian noise distributions, aperiodic signals, and biologically more realistic soft-threshold systems, thereby considerably broadening its potential relevance for neural information processing \cite{krauss2017adaptive}.

Building on this general framework, subsequent work extended the SR model beyond tinnitus-specific mechanisms and toward a broader theory of adaptive sensory processing and phantom perception \cite{krauss2018analysis, krauss2018cross, schilling2021stochastic}. The central idea was that internally generated neural noise constitutes a functional resource that can be dynamically regulated to maintain information transmission under degraded sensory conditions. Within this perspective, tinnitus and related phantom percepts emerge as side effects of an otherwise beneficial optimization process that improves signal detectability and sensory performance. This broader interpretation culminated in the concept of cross-modal stochastic resonance, which proposed that internally generated or cross-modal neural activity can serve as a functional source of noise that enhances sensory processing across modalities \cite{krauss2018cross}.

It was proposed that phantom percepts arise when adaptive sensory circuits attempt to maintain information transmission under persistently degraded input conditions. In this context, the model provided a mechanistic foundation for Fan Gang Zeng’s 2013 central noise model \cite{zeng2013active}. Thus, psychophysical findings have shown that central gain alone cannot explain tinnitus because tinnitus is not reliably associated with recruitment or hyperacusis or simple amplification of auditory stimuli \cite{zeng2013active}, and it often occurs without exaggerated sound-evoked responses. While the central noise model suggested a role for internally generated neural noise, it did not specify its origin. The stochastic resonance framework addressed this gap by identifying adaptive internal noise regulation as a concrete, circuit-level mechanism that enhances signal detection while generating phantom percepts as a byproduct \cite{schilling2022tinnitus, krauss2021simulated, zeng2013active, zeng2020tinnitus}.

In addition, the model evolved from a single-channel formulation to a multichannel perspective, in which frequency-specific pathways operate as partially independent adaptive detectors \cite{schilling2021stochastic}. Local regulation of internal noise in frequency regions affected by input degradation provides a mechanistic explanation for the frequency-specific characteristics of tinnitus and its clustering in regions associated with hearing loss. Subsequent work has related the framework to empirical findings, such as improved tone detection in external noise, transient phantom perceptions after sensory deprivation, somatosensory modulation of tinnitus, and the co-occurrence of tinnitus and hyperacusis \cite{shore2007neural, cederroth2020association}.

In 2021, evidence from an animal model showed that simulated transient hearing loss induced by long-term notched-noise exposure improved frequency-specific auditory sensitivity while simultaneously eliciting tinnitus-like phantom percepts, supporting the hypothesis that adaptive stochastic resonance enhances signal detection at the cost of internally generated phantom sounds \cite{krauss2021simulated}.

A particularly important milestone was the demonstration that stochastic resonance can improve complex auditory processing tasks beyond simple signal detection. Using a hybrid deep neural network model of the auditory pathway, Schilling and colleagues simulated the effects of hearing loss on speech recognition and investigated whether adaptive internal noise could compensate for degraded sensory input \cite{schilling2020intrinsic, schilling2022intrinsic}. Remarkably, the addition of intrinsic noise substantially improved speech recognition performance after simulated hearing loss, partially restoring performance levels that would otherwise have been lost. The results provided direct computational evidence that internally generated noise can serve a beneficial functional role in sensory systems by enhancing information transmission under degraded input conditions. Beyond its relevance for tinnitus, this study demonstrated that stochastic resonance may constitute a general computational principle for maintaining perceptual performance in impaired sensory systems.

Such functional benefits may help explain why some tinnitus patients exhibit less cognitive decline than expected given their degree of hearing loss \cite{hamza2021tinnitus, schilling2022tinnitus}.

Taken together, these developments support the idea that tinnitus is an emergent consequence of adaptive sensory optimization processes that aim to maintain information transmission under adverse sensory conditions.

\begin{landscape}
\begin{figure}[h]
    \centering
    \includegraphics[width=0.8\linewidth]{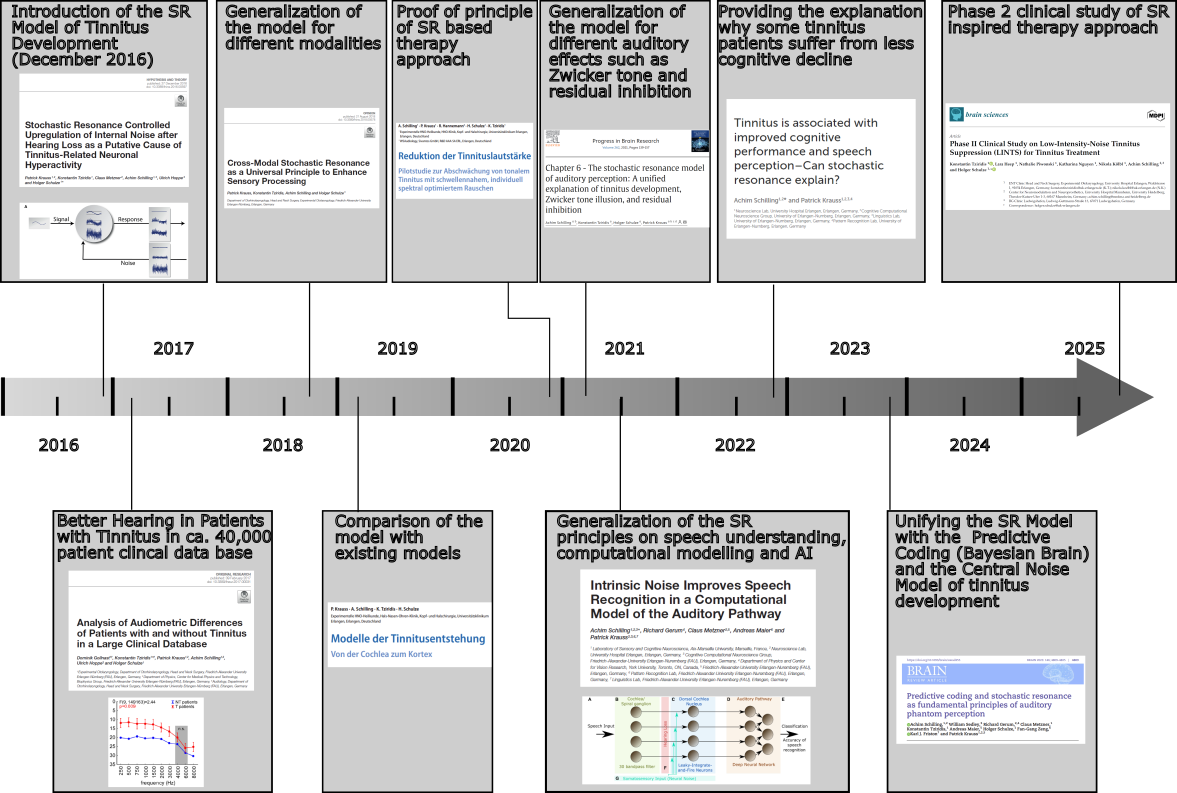}
    \caption{Timeline illustrating the development of the stochastic resonance (SR) model of tinnitus development from its initial formulation as a mechanism of tinnitus induction in the dorsal cochlear nucleus to a broader framework of adaptive sensory processing. Key milestones are summarized, including the first experimental evidence linking SR to tinnitus-related hyperactivity and better hearing thresholds; extensions to other sensory modalities and general principles of phantom perception; neural network simulations demonstrating benefits for speech processing and understanding; the development of an SR-based therapeutic approach; and, finally, the integration of SR with central gain and predictive coding theories into an integrated  model of tinnitus development. Figures taken from \cite{krauss2016stochastic, gollnast2017analysis, krauss2018cross, krauss2019models, schilling2020reduktion, schilling2022intrinsic, schilling2022tinnitus,schilling2023predictive,tziridis2025phase}.}
    \label{fig:history}
\end{figure}
\end{landscape}

\subsubsection{SR Model Based Therapy Approach}
The stochastic resonance framework was translated into a therapeutic strategy based on the central assumption that tinnitus typically emerges in frequency regions with the greatest hearing loss \cite{schecklmann2012relationship}. The core idea is to present narrow-band noise near the threshold, centered at the individual's tinnitus frequency. Introducing controlled external noise at the affected frequency effectively replaces or reduces internally generated neural noise. Functionally, this provides the auditory system with input in the previously deprived frequency band. This reduces the need for adaptive internal noise upregulation and attenuates the phantom percept \cite{schilling2020reduktion, tziridis2022spectrally, tziridis2025phase}.
Two pilot studies were conducted in 2021 and 2022 to evaluate the feasibility and short-term effects of this approach \cite{schilling2020reduktion, tziridis2022spectrally}. The results of these studies motivated a larger phase two clinical study, which was published at the end of 2025 \cite{tziridis2025phase}. This study demonstrated that the intervention is effective and minimally invasive. Notably, chronic improvements were also observed beyond immediate perceptual effects, suggesting that repeated external noise stimulation may induce longer-lasting adaptations. These findings provide initial clinical support for translating the stochastic resonance principle from a theoretical framework into a therapeutic application. However, further controlled trials are needed to determine the long-term efficacy of the intervention and the optimal implementation parameters.

\subsection{2023--now: Towards a Unified Theory of Tinnitus Development}

By 2023, it had become clear that no single model could account for the full range of tinnitus phenomena. Conceptual parallels had already been noted between the stochastic resonance framework and predictive or efficient coding theories, as both assume that sensory systems actively compensate for reduced or unreliable input to preserve stable representations \cite{sedley2016integrative, hullfish2019prediction, de2024bayesian}. However, these mechanisms operate at different explanatory levels. Stochastic resonance acts at the level of signal detection, modulating internal variability, and explaining the induction of tinnitus-related hyperactivity, particularly at the level of the dorsal cochlear nucleus \cite{schilling2023predictive}. Homeostatic plasticity and central gain mechanisms along the auditory pathway can subsequently amplify this hyperactivity, providing a link to the strength of recruitment and the frequent association between tinnitus and hyperacusis \cite{schilling2023predictive}. In contrast, predictive coding or Bayesian brain accounts address how such internally generated activity becomes a conscious perception. Within this unified perspective, stochastic resonance-induced hyperactivity may initially be misinterpreted as a real external tone, leading to acute tinnitus \cite{schilling2023predictive}. With ongoing internally generated noise, higher-level predictions gradually update, stabilizing the phantom perception and contributing to chronic tinnitus. In response to these complementary strengths, a consortium was formed in 2023 to integrate stochastic resonance, gain-based plasticity, and predictive coding into a generalized framework of tinnitus development. This framework connects the induction, amplification, and conscious perception of tinnitus within a single, coherent model \cite{schilling2023predictive}.

\begin{figure}[h!]
    \centering
    \includegraphics[width=0.6\linewidth]{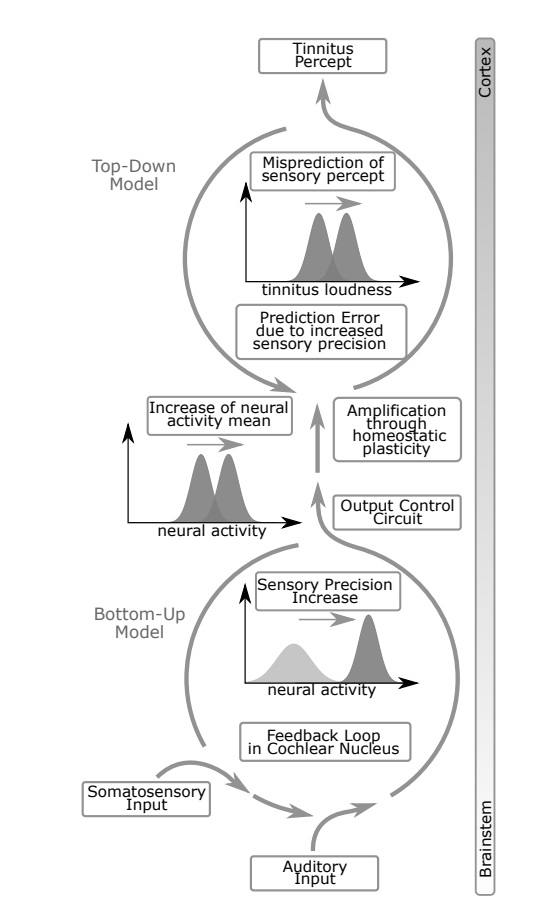}
    \caption{\small{Integrated model of tinnitus development that combines SR, central gain and predictive coding mechanisms. Initially, reduced auditory input triggers adaptive internal noise regulation within the DCN via a local feedback loop that optimizes signal detectability by adjusting the amount of internally generated noise. This SR-related hyperactivity can induce acute tinnitus-related activity patterns. Along the auditory pathway, homeostatic plasticity and central gain mechanisms further amplify this neural activity, inducing hyperacusis and increasing tinnitus strength. At higher processing stages, including thalamic and cortical regions, a second feedback loop minimizes prediction error by integrating internally generated activity into perceptual inference. Initially, the stochastic resonance-induced activity may be misinterpreted as an external tone, leading to the conscious perception of tinnitus. Over time, persistent internally generated activity updates higher-level predictions, stabilizing and chronicling the phantom perception.}}
    \label{fig:model}
\end{figure}

\section{The Future of the SR Model and Tinnitus Research}
Ten years after its introduction, the stochastic resonance framework has evolved from a theoretical proposal into a unifying perspective on tinnitus development, connecting auditory physiology, computational neuroscience, and adaptive information processing. By interpreting tinnitus-related hyperactivity as a consequence of sensory optimization rather than pathological overexcitation, the model shifts the focus from malfunction to adaptation. It also reframes phantom perception as a potential byproduct of mechanisms that normally preserve sensitivity under degraded input conditions.

The next decisive step is empirical verification at the neural level. Demonstrating adaptive internal noise regulation directly within the auditory system would provide critical support for the framework. Such evidence would elevate the model from theoretical plausibility to mechanistic validation. In parallel, rigorous quantitative testing is required. The stochastic resonance framework makes concrete, falsifiable predictions, such as improved detectability under specific noise regimes and the coexistence of enhanced sensitivity with phantom perceptions. Experimental paradigms capable of distinguishing these dynamics from explanations based solely on gain changes or plasticity are essential.

Clinical translation further raises the stakes. The initial therapeutic applications derived from the model suggest that adaptive noise modulation is not only theoretically significant, but also clinically relevant. However, large controlled trials and objective biomarkers are now required to define responder profiles and optimize treatment parameters.

Importantly, the stochastic resonance framework does not oppose existing models, but rather, it integrates induction, amplification, and conscious perception within a broader adaptive architecture. Whether stochastic resonance is the primary driver of tinnitus or represents a core mechanism within a multilevel process remains to be determined empirically. However, it is clear that the model has redefined tinnitus, shifting its perception from a purely maladaptive phenomenon to a potentially emergent property of adaptive sensory computation. This conceptual shift opens a new research program linking basic neuroscience, computational modeling, and clinical intervention within a coherent, testable framework.


\section*{Author contributions}
Both authors contributed equally to this article.

\section*{Acknowledgments}
The authors would like to thank Arnaud Norena, Holger Schulze, and Konstantin Tziridis for many stimulating discussions and valuable scientific exchanges that have contributed to the development of the stochastic resonance framework over the past decade. We are also grateful to the numerous anonymous reviewers of our manuscripts and grant proposals whose constructive criticism, suggestions, and thoughtful comments helped refine both the theoretical foundations and empirical investigations of the model. Their feedback has played an important role in shaping the work summarized in this review.

\section*{Funding}
This work was funded by the Deutsche Forschungsgemeinschaft (DFG, German Research Foundation): grants KR\,5148/3-1 (project number 510395418), KR\,5148/5-1 (project number 542747151), KR\,5148/10-1 (project number 563909707) and GRK\,2839 (project number 468527017) to PK, and grants SCHI\,1482/3-1 (project number 451810794) and SCHI\,1482/6-1 (project number 563909707) to AS.



\bibliographystyle{apalike}
\bibliography{references}

\end{document}